# Thing-to-Thing Optical Wireless Power Transfer Based on Metal Halide Perovskite Transceivers

Dinh Hoa Nguyen[†,1,2], Chuanjiang Qin,[3–5] Toshinori Matsushima[1,5], Chihaya Adachi[1,4,5]

[1]International Institute for Carbon-Neutral Energy Research (WPI-I2CNER), Kyushu University, 744 Motooka, Nishi, Fukuoka 819-0395, Japan

[2]Institute of Mathematics for Industry (IMI), Kyushu University, 744 Motooka, Nishi, Fukuoka 819-0395, Japan

[3]State Key Laboratory of Polymer Physics and Chemistry, Changchun Institute of Applied Chemistry (CIAC), Chinese Academy of Science, 5625 Renmin Street, Changchun 130022, China

[4]Center for Organic Photonics and Electronics Research (OPERA), Kyushu University, 744 Motooka, Nishi, Fukuoka 819-0395, Japan

[5]Japan Science and Technology Agency (JST), ERATO, Adachi Molecular Exciton Engineering Project, 744 Motooka, Nishi, Fukuoka 819-0395, Japan

**Abstract.** This paper proposes a novel conceptual system of optical wireless power transfer between objects, whether stationary or in-motion. Different from the currently existing optical wireless power transfer systems, where the optical transmitter and receiver are two distinct devices, the proposed system in this paper employs a single device – an optical transceiver, which is capable of working as both light absorption and emission. This optical transceiver is fabricated from a metal halide perovskite which yields superior features such as low costs, capability of flexibly attached on curved surfaces, thin and light weight. Therefore, the whole system size and cost can be significantly reduced, while perovskite transceivers can be made adaptive to any surface. This will contribute to realize a thing-to-thing optical wireless power transfer system, in which surfaces of objects/things are covered (fully or partially) by perovskite transceivers, enabling them to wirelessly charge or discharge from the others.

## 1. Introduction

Wireless power transfer (WPT) has recently received much attention across many disciplines in both academic and industrial research and applications because of its numerous advantages over wired power transfer techniques. For instance, WPT is much more convenient and comfortable for users in daily life, where users do not need to take the wires for plugging into their electric vehicles (EVs), but instead just need to park their EVs over wireless charging pads. Also, fire and explosion problems with wired charging can be avoided. Another benefit of WPT is that it is indispensable for many circumstances where wired power transfer is impossible, difficult, expensive, or dangerous such as with high-voltage power applications [1], space [2], [14] or deep water exploration missions.

[†]Corresponding author. Email: hoa.nd@i2cner.kyushu-u.ac.jp

Several methods of WPT have been proposed including inductive WPT (IWPT) and resonant IWPT [3]–[6], capacitive WPT (CWPT) [7], and optical WPT (OWPT) [8]–[12] for different ground, aerial, and underwater applications. It can be seen from the above research works that the highest WPT efficiency is currently achieved with resonant IWPT technologies, however the transfer distance is limited. The energy efficiency of CWPT technology is now closer to that of the resonant IWPT technology, but CWPT is harmful for people around the CWPT devices. To extend transfer distance while keeping the related system components rather simple, OWPT is one of the most suitable approaches. A few companies, e.g., [10], have been investigating this technology for consumer electronics and other fields. Meanwhile, several research institutes and agencies around the world are investigating OWPT for far-field WPT, e.g. [14], [15], [16]. It is noted that laser is dominant in current OWPT research (see, e.g. [10], [11], [12]), however it is not safe for human, especially when exposed to eyes. On the other hand, light-emitting diodes (LEDs) are safer to be utilized in OWPT systems, but the generated light from them are more diverse than that from laser beams, leading to decreased efficiency. Hence, there is always a tradeoff between efficiency and safety in OWPT systems when using lasers and LEDs.

It is worth emphasizing that in all of the aforementioned studies on WPT, at least the wireless energy transmitter, if not both the energy transmitter and the energy receiver, is stationary. For example, the recent concept of wireless charging lane (see, e.g. [17]) allows dynamic charging of EVs while they are moving, but the under-road charging coils of wireless charging lane are fixed. Similarly, the system in [13] is able to wirelessly charge EVs when they are moving around, nevertheless the optical energy transmitter is fixed on the ceiling. Those concept and system are great in the sense that they do not require wireless energy receivers to be standstill, but the energy transmitters need to be fixed. Hence, the energy mobility and user convenience are not fully exploited.

The current research is built upon the existing WPT technologies to extend them for the bidirectional WPT, in particular the bidirectional OWPT, between different objects, whether stationary or in-motion. To achieve that, we propose a novel OWPT system, in which a *metal halide perovskite transceiver* is employed for both absorbing light from and emitting light to other OWPT systems equipped with similar perovskite transceivers. As such, the limitation on unidirectional energy transfer of the current OWPT system is overcame. Further, our proposed OWPT system enables wireless energy transfer between objects that are all moving, because such system can be easily switched from energy reception mode to energy transmission mode anywhere, at any time. Therefore, our proposed perovskite-transceiver-based OWPT system hurdles the limitations of current WPT systems and underpins the ultra-mobility and convenience for energy transfer.

The rest of this paper is organized as follows. Section 2 introduces our proposed bidirectional OWPT using perovskite transceivers for realization of a thing-to-thing (T2T) OWPT network, and remarkable properties of perovskite-based devices to be used as optical transceivers in the proposed OWPT system, supported by experimental results. Then the system efficiency is theoretically characterized in Section 3. Lastly, Section 4 concludes our study, and provide discussions on the usefulness of the proposed system, as well as directions for future research.

## 2. T2T-OWPT based on Perovskite Transceiver

### 2.1. Illustrative System Configuration

Figure 1 depicts the proposed OWPT system composing of basic elements such as devices whose surfaces are covered with optical transceivers and a transmission environment. The optical transceiver is made from a metal halide perovskite film whose properties will be introduced in the next section. The transmission environment could be air, water, sea water, or outer space, depending on specific situations on which WPT is needed. Having the nice features of bidirectional transfer and ultra-mobile, the proposed OWPT system helps realize the so-called T2T-OWPT network, in which energy can be ubiquitously exchanged anywhere, anytime between multiple objects/things equipped with perovskite transceivers, whether stationary or moving.

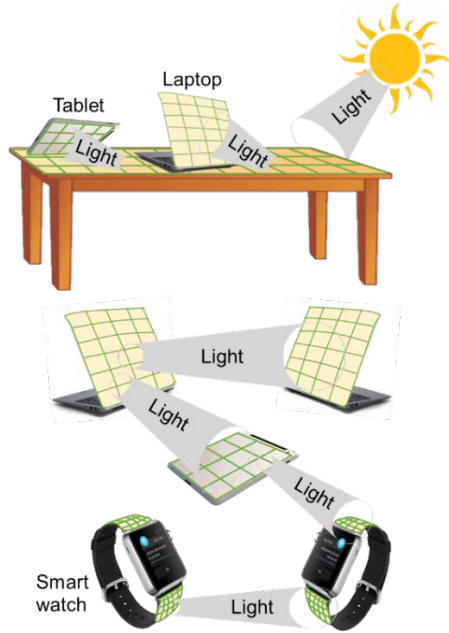

Figure 1. Illustration of the proposed conceptual system of T2T-OWPT.

## 2.2. Properties of Perovskite Transceiver

Metal halide perovskites have many advantages such as high carrier mobilities, large carrier diffusion lengths, high absorption coefficients, bandgap tunability, and compatibility with low-cost solution processing. These advantages make perovskites suitable for use in solar cells. Since the first discovery of perovskite solar cells in 2009 [18], their performance has rapidly been increased. The highest certified power conversion efficiency of perovskite solar cells has reached 25.2% [19] which is comparable to silicon solar cell technology. Additionally, by taking advantage of efficient photoluminescence, perovskites can be employed as the emitter of efficient LEDs. Recently, very high external quantum efficiencies of over 20% have been demonstrated from perovskite LEDs [20–22].

Our group in Kyushu University has investigated perovskite solar cells to improve their power conversion efficiencies and operational durability, e.g. [23–27]. Our perovskite solar cells reported in [27] had an architecture of glass substrate / indium tin oxide (ITO) electrode / poly(3,4-ethylenedioxythiophene):poly-(styrenesulfonate) (PEDOT:PSS) hole transport layer / $MA_{0.6}FA_{0.4}PbI_{2.8}Br_{0.2}$ perovskite light absorber (MA is methylammonium and FA is formamidinium) / $C_{60}$ electron transport layer / 2,9-dimethyl-4,7- diphenyl-1,10-phenanthroline (BCP) electron transport layer / Au electrode, and exhibited a power conversion efficiency of ~15%. Detailed experimental results and conditions can be found in [27]. When we applied a high voltage to the same perovskite solar cells, electrons and holes were injected from electrodes and recombined in the perovskite layer, which is a completely opposite process compared with solar cells. As a result, the perovskite solar cells began to emit light. Figure 2(a) displays a photograph of our perovskite solar cell emitting near-infrared (NIR) light under the electrical excitation. These devices exhibited an electroluminescence (EL) peak at ~790 nm as shown in Figure 2(b). Emission at longer wavelengths is possible by utilizing smaller-bandgap perovskites such as tin iodide-based perovskites [28]. This means that perovskite solar cells can potentially work as LEDs. Therefore, we can use perovskites for both the light absorber of solar cells (optical receivers for OWPT) and the emitter of LEDs (optical transmitters for OWPT) simultaneously in a single-device architecture, i.e., an optical transceiver. Other merit of using perovskite LEDs is efficient emission of NIR light [22], which is useful for OWPT because NIR light is invisible and not so harmful for human. Additionally, perovskites are known to have the large spectral overlap between the emission and the absorption. Incident photon-to-electron conversion efficiencies (IPCEs) of the aforementioned perovskite solar cells were measured and plotted as a function of wavelength in Figure 2(b). It is clear that the EL peak overlaps with the IPCE curve, pointing to the self-absorption of light by the perovskite. In other words, perovskites show light emission near the band edge. These features make it possible to realize that one device can receive optical power emitted from the other device even with the same perovskite composition and device architecture.

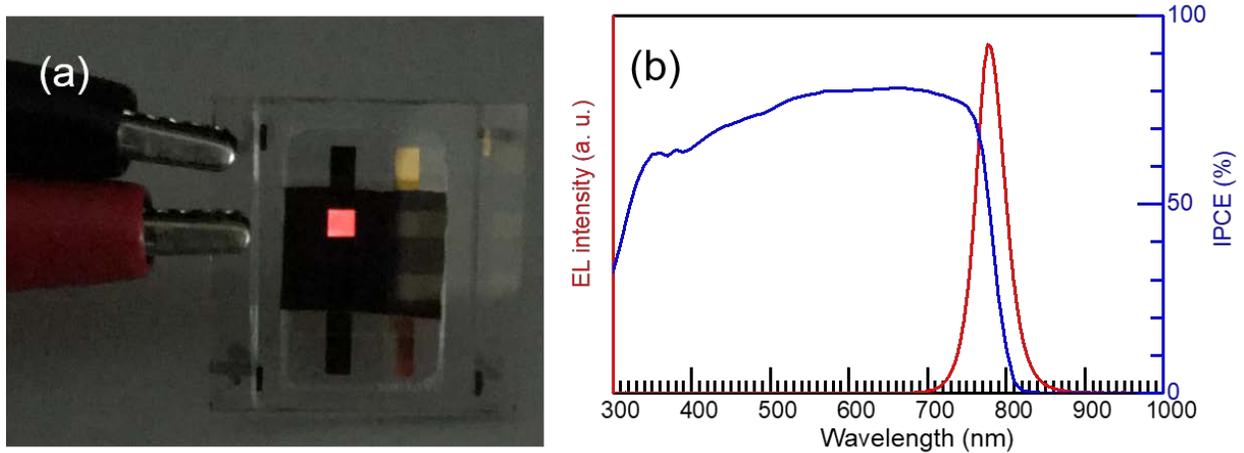

Figure 2. (a) Photograph of an NIR light-emitting perovskite solar cell. (b) EL spectrum and IPCE curve of the same device.

## 3. Efficiency of The Proposed T2T-OWPT System

### 3.1. Computation of Whole System Efficiency

The OWPT system efficiency denoted by $\eta_{sys}$ is derived by

$$\eta_{sys} = \eta_{tr} \times \eta_{env} \times \eta_{rc} \tag{1}$$

where $\eta_{tr}$, $\eta_{env}$, $\eta_{rc}$ represent the efficiency of the optical transmitter, of the optical transfer through the environment, and of the optical receiver; respectively. This formula can be used to explain the overall energy conversion efficiency of an OWPT system.

In eq. (1), $\eta_{tr}$ represents the energy efficiency of the optical transmitter, which is the ratio of its output energy to its consumed energy. Next, $\eta_{env}$ shows how much energy is lost when light is traveling in the considering environment such as air, pure water or sea water, etc. Lastly, $\eta_{rc}$ describes the efficiency of the optical energy receiver when converting light energy to electric energy. In the following mathematical formula for the efficiency $\eta_{env}$ of the optical transmission through the environment is presented. For simplicity, it is assumed that light from one perovskite device is transmitted through the air to another perovskite device. Optical transmission efficiency in other environments, e.g., pure water, sea water, etc., can be derived similarly.

The transmitted optical power (or radiant flux) from a perovskite device working as an LED is computed by (see, e.g. [29]),

$$P_{tr} = \int_{\lambda_{min}}^{\lambda_{max}} \int_0^{2\pi} \Phi_e(\theta, \lambda) d\theta d\lambda \quad (2)$$

which is the integral of the energy flux $\Phi_e$ in all directions. In equation (2), $\lambda_{min}$ and $\lambda_{max}$ are minimum and maximum wavelength can be detected by the optical receiver, i.e. the receiving perovskite solar cell.

Next, the received optical power at the receiver surface, assumed direct line of sight transmission, is computed as follows [29],

$$P_{rc} = H(0)P_{tr} \quad (3)$$

where $H(0)$ is the DC gain of the optical transmission link calculated by

$$H(0) = \begin{cases} \frac{(m_l+1)A}{2\pi d^2} \cos^{m_l} \phi \, g_f(\varphi) g_c(\varphi) \cos \varphi : 0 \leq \varphi \leq \varphi_w \\ 0 : \varphi > \varphi_w \end{cases} \quad (4)$$

in which $A$ is the physical area of the optical receiver, $d$ is the optical transmission distance, $\varphi$ is the angle of incidence (i.e. the angle at which the receiver sees the transmitter), $\phi$ is the angle of irradiance (i.e. the angle at which the transmitter sees the receiver), $g_f(\varphi)$ is the gain of an optical filter if exists, $g_c(\varphi)$ is the gain of an optical concentrator if exists, and $\varphi_w$ is the width of the field of view (FOV) at the optical receiver. Moreover, $m_l$ denotes the order of Lambertian emission which is given by the semi-angle at half illuminance $\phi_{1/2}$ of an LED as follows,

$$m_l = -\frac{\ln 2}{\ln(\cos \phi_{1/2})}$$

Here, $H(0)$ represents a mathematical computation of $\eta_{env}$. It is obviously from eq. (4) that the transmitted optical power is inversely quadratically decreased with the transmission distance, when the angle of incidence is not greater than the angle of irradiance.

### 3.2. Efficiencies of Perovskite Transceiver

The experimental IPCE value was ~46% at the EL peak wavelength [Figure 2(b)]. The external quantum (photon-to-electron conversion) efficiency of the aforementioned light-emitting perovskite solar cells was still low at ~1%. We need a further study to increase the external quantum efficiency in the future. We tentatively took these values as the efficiencies of the optical receiver ($\eta_{rc}$) and transmitter ($\eta_{tr}$) to estimate the OWPT system efficiency. Thus, the efficiency of the whole bidirectional OWPT system based on metal halide perovskite transceiver, at zero transmission distance, is 0.46% at present. The highest external quantum efficiency reported from NIR perovskite LEDs to date was 21.6% [22]. Using this highest value leads to an increase in OWPT system efficiency to 9.94%.

## 4. Conclusion and Discussion

### 4.1. Summary

This article has proposed a novel conceptual OWPT system of between objects equipped with optical transceivers made from metal halide perovskite devices. As the results, each of such objects can work as both a light absorber and a light emitter, leading to bidirectional OWPT between those objects, whether they are stationary or moving. Theoretical characterization of system energy conversion efficiency has been provided, and experimental results on the light absorption and emission modes of perovskite devices have been reported.

Based on the experimental results and those reported in the literature, system efficiency of the proposed perovskite-transceiver-based OWPT system is now 9.94% at maximum, which is low compared to that of the resonant IWPT system. However, the payoffs of rather simple electronic requirement, potentially low cost, usability for curved surfaces, printability, and full support for moving objects render the proposed OWPT system appropriate for ubiquitous applications, especially in the current internet-of-thing (IoT) era. Further, the proposed system helps realize a T2T-OWPT network, in which OWPT can be made anywhere, anytime between diverse entities (things), e.g., consumer electronic devices, implant medical devices, robots, IoT devices, etc. Such T2T-OWPT network will bring not only the ultra-mobility of energy but also the convenience and comfort for human.

### 4.2. Further Discussions

In the area of vehicle electrification, our proposed OWPT system could impinge upon novel wireless charging and discharging methods for electrified vehicles covered with perovskite solar cells. It is worth mentioning that solar-cell-covered vehicles have recently been initiated by several automobile companies, e.g., [30], [31], [32].

It is also worth noting that laser can be used in the proposed perovskite-transceiver-based bidirectional OWPT system, in which laser is generated in the light emitting mode instead of LED light. Due to the coherence of laser beams, higher overall system energy conversion efficiency and longer WPT range could be achieved. Although laser is harmful for human if the energy level is high, as discussed before, this kind of system could be useful in specific applications where human and other species are not present, and energy grids do not exist, e.g., in space exploration missions (for instance the Moon or Mars exploration). In such applications, solar energy is harvested by spacecrafts, satellites, landers, rovers, etc., equipped with the proposed bidirectional perovskite-transceiver-based OWPT systems, and then is wirelessly transmitted to other spacecrafts, satellites, landers, rovers, etc., when they lack energy but cannot receive it from solar. As a result, mission period can be extended smoothly, which is very important for such space exploration applications. Recent progresses in perovskite laser research (see, e.g., reviews in [33], [34], [35], and our recent work [36]) suggest that electrically driven laser based on metal halide perovskite is promising, hence can be employed for our proposed perovskite-transceiver-based bidirectional OWPT system.

**Acknowledgement**

The first author's research is partially funded by Horiba Ltd. through the Masao Horiba Awards. Additionally, this research was supported by the Japan Science and Technology Agency (JST), ERATO, Adachi Molecular Exciton Engineering Project (JST ERATO Grant Number JPMJER1305); JSPS KAKENHI (grant numbers JP16H04192 and 20H02817); The Canon Foundation; and The Samco Foundation.

**Conflict of Interest**

The authors declare no conflict of interest.